\documentstyle[12pt]{article}

\font\blackboard=msbm10 at 12pt
\font\blackboards=msbm7
\font\blackboardss=msbm5
\newfam\black
\textfont\black=\blackboard
\scriptfont\black=\blackboards
\scriptscriptfont\black=\blackboardss

\def\bfx{{\bf X}}
\def\bfy{{\bf Y}}

\newcommand{\ba}{\begin{array}}
\newcommand{\ea}{\end{array}}
\newcommand{\be}{\begin{equation}}
\newcommand{\ee}{\end{equation}}
\newcommand{\bea}{\begin{eqnarray}}
\newcommand{\eea}{\end{eqnarray}}
\newcommand{\beas}{\begin{eqnarray*}}
\newcommand{\eeas}{\end{eqnarray*}}

\def\half{{1 \over 2}}
\def\identity{{\rlap{1} \hskip 1.6pt \hbox{1}}}

\def\laplace{{\kern1pt\vbox{\hrule height 1.2pt\hbox{\vrule width 1.2pt\hskip
  3pt\vbox{\vskip 6pt}\hskip 3pt\vrule width 0.6pt}\hrule height 0.6pt}
  \kern1pt}}
\def\scriptlap{{\kern1pt\vbox{\hrule height 0.8pt\hbox{\vrule width 0.8pt
  \hskip2pt\vbox{\vskip 4pt}\hskip 2pt\vrule width 0.4pt}\hrule height 0.4pt}
  \kern1pt}}

\def\roughly#1{\raise.3ex\hbox{$#1$\kern-.75em\lower1ex\hbox{$\sim$}}}

\def\gym{g^2_{\scriptscriptstyle YM}}

\def\tr{{\rm Tr} \,}
\def\trl{{\rm Tr}_L}
\def\trg{{\rm Tr}_G}

\newcommand{\NP}{{\em Nucl.\ Phys.\ }}
\newcommand{\PL}{{\em Phys.\ Lett.\ }}

\newcommand{\PRL}{{\em Phys.\ Rev.\ Lett.\ }}

\newcommand{\gone}[1]{}
\begin{document}
\pagestyle{plain}
\setcounter{page}{1}

\baselineskip16pt

\begin{titlepage}

\begin{flushright}
IASSNS-HEP-97/121\\
PUPT-1734\\
hep-th/9711078
\end{flushright}
\vspace{16 mm}

\begin{center}
{\Large \bf Spherical membranes in Matrix theory}

\vspace{3mm}

\end{center}

\vspace{8 mm}

\begin{center}

Daniel Kabat${}^1$ and Washington Taylor IV${}^2$

\vspace{3mm}

${}^1${\small \sl School of Natural Sciences} \\
{\small \sl Institute for Advanced Study} \\
{\small \sl Olden Lane} \\
{\small \sl Princeton, New Jersey 08540, U.S.A.} \\
{\small \tt kabat@ias.edu}

\vspace{3mm}

${}^2${\small \sl Department of Physics} \\
{\small \sl Joseph Henry Laboratories} \\
{\small \sl Princeton University} \\
{\small \sl Princeton, New Jersey 08544, U.S.A.} \\
{\small \tt wati@princeton.edu}

\end{center}

\vspace{8mm}

\begin{abstract}
We consider membranes of spherical topology in uncompactified Matrix
theory.  In general for large membranes Matrix theory reproduces the
classical membrane dynamics up to $1/N$ corrections; for certain
simple membrane configurations, the equations of motion agree exactly
at finite $N$.  We derive a general formula for the one-loop Matrix
potential between two finite-sized objects at large separations.
Applied to a graviton interacting with a round spherical membrane, we
show that the Matrix potential agrees with the naive supergravity
potential for large $N$, but differs at subleading orders in $N$.  The
result is quite general: we prove a pair of theorems showing that for
large $N$, after removing the effects of gravitational radiation, the
one-loop potential between classical Matrix configurations agrees
with the long-distance potential expected from supergravity.  As a
spherical membrane shrinks, it eventually becomes a black hole.  This
provides a natural framework to study Schwarzschild black holes in
Matrix theory.
\end{abstract}

\vspace{8mm}
\begin{flushleft}
November 1997
\end{flushleft}
\end{titlepage}
\newpage

\section{Introduction}
\label{Intro}

It was conjectured by Banks, Fischler, Shenker, and Susskind (BFSS)
that M-theory in the infinite momentum frame is exactly described by a
supersymmetric matrix quantum mechanics \cite{BFSS}.  This seminal
conjecture has passed many consistency checks.  Gravitons \cite{BFSS},
membranes \cite{Goldstone-Hoppe,dhn}, and 5-branes
\cite{Berkooz-Douglas,grt,bss} have been constructed as states in
Matrix theory corresponding to block matrices with particular
commutation properties.  Numerous calculations have shown that at
one-loop order Matrix theory correctly reproduces long-range forces
expected from supergravity (see for example
\cite{BFSS,DKPS,Aharony-Berkooz,Lifschytz-Mathur,Lifschytz-46,bc,ChepTseyI,MaldacenaI,Vakkuri-Kraus,Gopakumar-Ramgoolam,ChepelevTseytlin,MaldacenaII}).
Some progress has been made in understanding two-loop corrections
\cite{Becker-Becker,ggr,bbpt,ChepelevTseytlin,Esko-Per2} as well as
processes involving M-momentum transfer \cite{pp}.  The conjecture has
been put on a more systematic footing by the work of Sen and Seiberg
\cite{Sen,Seiberg-DLCQ}.  For a recent review see \cite{banks-review}.

To date, most discussions of extended objects in Matrix theory have
focused on static branes, which are either infinite in extent or have
been wrapped around toroidally compactified dimensions.  Such
membranes preserve some supersymmetry.  In this paper, we focus on a
different class of states, namely membranes of spherical topology.
Such membranes break all of the supersymmetries of the theory, and as
solutions to the equations of motion are time-dependent.  Nonetheless,
the means for studying such objects was developed long ago
\cite{Goldstone-Hoppe,dhn}.  In light-front gauge there is an explicit
map between embeddings of a spherical membrane and Hermitian matrices.
In this paper, we use this map to discuss spherical membranes in
uncompactified Matrix theory.  Although we focus here on membranes
with spherical topology, analogous results could be derived for higher
genus membranes \cite{Hoppe-Nicolai,Hoppe}.

Our calculation of the Matrix theory potential will be performed for
finite-sized matrices, and thus should be understood in the context of
discrete light-front quantization \cite{Susskind-DLCQ}.  We will see
that the Matrix potential only agrees with the naive discrete
light-front quantization of supergravity at leading order in large
$N$, while the subleading terms in $N$ differ.  This supports the
conjecture that large-$N$ Matrix theory is equivalent to
11-dimensional M-theory.  But it also supports the idea that
finite-$N$ Matrix theory, although equivalent to the DLCQ of M-theory
\cite{Seiberg-DLCQ}, simply does not reduce to the DLCQ of
supergravity at low energies \cite{banks-review,hp}.  Other
finite-$N$ discrepancies have been found, involving curved backgrounds
in Matrix theory \cite{dos,DineRajaraman,DouglasOoguri}.

Our primary motivation for this paper is simply to understand the
behavior of large, semiclassical membranes in Matrix theory.  But
ultimately a more interesting set of questions involves the behavior
of a membrane in the quantum domain.  A spherical membrane will shrink
with time until it is smaller than its Schwarzschild radius.
Eventually quantum effects will become important, and cause the
individual partons composing the membrane to fly off in different
directions.  Thus spherical membranes provide a natural context for
studying black holes in Matrix theory.

The structure of this paper is as follows.  In Section 2, we review
the prescription developed in \cite{dhn} for describing spherical
membranes in Matrix theory.  We show that even at finite $N$ one has
the expected interpretation in terms of a spherical geometry.  We also
show that for large membranes, Matrix theory reproduces the
semiclassical membrane equations, up to $1/N$ corrections.  In Section
3, we derive a general formula for the one-loop Matrix theory
potential between two finite-sized objects.  In Section 4, we compare
the Matrix calculation to the results expected from supergravity.  We
show that the potential between a spherical membrane and a graviton is
correctly reproduced at leading order in $N$ but that there are
discrepancies at subleading order.  We also prove a general pair of
theorems which show that, after removing the effects of gravitational
radiation, the leading term in the supergravity potential is always
reproduced by large $N$ Matrix theory.  Section 5 contains
conclusions and a discussion of further directions for research,
including some comments about Matrix black holes.

\section{Matrix description of spherical membranes}
\label{sec:description}

\subsection{Membrane-matrix correspondence}

In the original work of de Wit, Hoppe and Nicolai \cite{dhn}, the
supermembrane was quantized in light front gauge.  An essential
observation of these authors was that the Lie algebra of
area-preserving diffeomorphisms of the sphere can be approximated
arbitrarily well by the algebra of $N \times N$ Hermitian matrices, when
$N$ is taken to be sufficiently large.  The Poisson bracket of two
functions on the sphere is then naturally associated with a commutator
in the $U(N)$ algebra.  This correspondence can be made explicit by
considering the three Cartesian coordinate functions $x_1, x_2, x_3$
on the unit sphere.  With a rotationally invariant area form (which we
identify with a symplectic form), these Cartesian coordinates are
generators whose Hamiltonian flows correspond to rotations around the
three axes of the sphere.  The Poisson brackets of these functions are
given by
\[
\{x_A, x_B\} = \epsilon_{ABC} x_C.
\]
It is therefore natural to associate these coordinate functions on $S^2$
with the generators of $SU(2)$.  The correct correspondence turns out
to be
\[
x_A \leftrightarrow {2 \over N} J_A
\]
where $J_1, J_2, J_3$ are generators of the $N$-dimensional representation of
$SU(2)$, satisfying the commutation relations
\[
-i [J_A, J_B] = \epsilon_{ABC} J_C \,.
\]

In general, the transverse coordinates of a membrane of spherical
topology can be expanded as a sum of polynomials
of the coordinate functions:
\[
X^i (x_A)= \sum_k t^i_{A_1 \ldots A_k} x_{A_1} \cdots x_{A_k}\,.
\]
The coefficients $t^i_{A_1 \ldots A_k}$ are symmetric and traceless
(because $x_A x_A = 1$).  Using the above correspondence, a matrix
approximation to this membrane can be constructed, which we denote by
${\bf X}^i$.  It is necessary to truncate the expansion at terms of
order $N-1$, because higher order terms in the power series do not
generate linearly independent matrices
\be
\label{MMcorrespond}
{\bf X}^i = 
\sum_{k < N} \left(2 \over N\right)^k t^i_{A_1 \ldots A_k} J_{A_1}
\cdots J_{A_k}.
\ee
This is the map between membrane configurations and states in Matrix
theory developed in \cite{dhn}.

To motivate the correspondence, we begin by recording the membrane
equations of motion in light-front gauge.  The membrane action
is\footnote{The spacetime metric is $\eta_{\mu\nu}=(-+\cdots+)$.
Light-front coordinates are $X^\pm \equiv {1 \over \sqrt{2}}\left(X^0
\pm X^{10}\right)$.  $X^+$ is light-front time; the conjugate energy
is $p^-={1 \over \sqrt{2}}\left(E-p^{10}\right)$.  The coordinate
$X^-$ is compactified, $X^- \approx X^- + 2 \pi R$, and the conjugate
momentum $p^+ = {1 \over \sqrt{2}}\left(E+p^{10}\right)$ is quantized,
$p^+ = N/R$.  The tension of the membrane is $T_2 = {1 \over (2 \pi)^2
l_P^3}$ (in the conventions of \cite{dealwis-normalize}).  We adopt
units which lead to a canonical Yang-Mills action: $2 \pi l_P^3 = R$.}
\[
S = - {T_2 \over 2} \int d^3 \xi \sqrt{-g} \left( g^{\alpha \beta} \partial_\alpha X^\mu \partial_\beta X^\nu
\eta_{\mu\nu} - 1 \right)\,.
\]
Here $\xi^\alpha = \left(t,\sigma^a\right)$ are coordinates on the
membrane worldvolume, $g_{\alpha\beta}$ is an auxiliary worldvolume
metric, and $X^\mu(\xi)$ are embedding coordinates.  The light-front
gauge fixing of this action is carried out in \cite{dhn,bst}.  In
terms of equal-area coordinates $\sigma^a$ on a round unit two-sphere,
the equations of motion can be written either in terms of an induced
metric $\gamma_{ab} \equiv \partial_a X^i \partial_b X^i$, or in terms
of a Poisson bracket $\{f,g\} \equiv \epsilon^{ab} \partial_a f
\partial_b g$ where $\epsilon^{12} = 1$.  In light-front gauge the fields
$X^+$, $X^-$ are constrained:
\bea
X^+ &=& t  \nonumber\\
\noalign{\vskip 0.2 cm}\label{constraints}
\dot{X}^- &=& \half \dot{X}^i \dot{X}^i + {2 \gamma \over N^2} \\
\noalign{\vskip 0.1 cm}
          &=& \half \dot{X}^i \dot{X}^i + {1 \over N^2}
\{X^i,X^j\}\{X^i,X^j\} \nonumber\\
\noalign{\vskip 0.2 cm}
\partial_a X^- &=& \dot{X}^i \partial_a X^i \nonumber
\eea
where $\gamma \equiv \det \gamma_{ab} = \half \{X^i,X^j\}\{X^i,X^j\}$.  The
transverse coordinates $X^i$ have dynamical equations of motion
\bea
\label{dynamical}
\ddot{X}^i &=& {4 \over N^2} \partial_a \left(\gamma \gamma^{ab}
\partial_b X^i\right) \nonumber \\
           &=& {4 \over N^2} \{\{X^i,X^j\},X^j\} \nonumber
\eea
which follow from the Hamiltonian
\bea
\label{MembraneHamiltonian}
H &=& {N \over 4 \pi R} \int d^2 \sigma \left( \half \dot{X}^i \dot{X}^i
+ {2 \gamma \over N^2} \right) \\
\noalign{\vskip 0.1 cm}
  &=& {N \over 4 \pi R} \int d^2 \sigma \left( \half \dot{X}^i \dot{X}^i
+ {1 \over N^2} \{X^i,X^j\}\{X^i,X^j\} \right)\,. \nonumber
\eea
The transverse coordinates must obey the condition
\be
\label{MembraneGauss}
\{\dot{X}^i,X^i\}=0 \ee so that the constraint on $\partial_a X^-$ can
be satisfied.  Note that the velocity in $X^-$ is essentially the
energy density on the membrane.  The density of momentum conjugate to
$X^-$ is constant on the membrane, and we have chosen a normalization
so that the total momentum takes on the expected value.
\[
p^+ = T_2 \int d^2 \sigma \sqrt{-\gamma/g_{00}} \,\,
\dot{X}^+ = {N \over R}
\]
Here $g_{00} \equiv \eta_{\mu\nu} \dot{X}^\mu \dot{X}^\nu$.  Then using
the matrix correspondence
\[
x_A \leftrightarrow {2 \over N} J_A \hskip 1.0 cm
\{\cdot,\cdot\} \leftrightarrow  {-i N \over 2} [\cdot,\cdot] \hskip 1.0 cm
{N \over 4 \pi} \int d^2 \sigma \leftrightarrow \tr
\]
we see that the membrane Hamiltonian becomes the (bosonic part of) the
Hamiltonian of Matrix theory
\be
\label{MatrixHamiltonian}
H = {1 \over R} \tr \left( \half \dot{{\bf X}}^i \dot{{\bf X}}^i - {1 \over 4}
[{\bf X}^i,{\bf X}^j] [{\bf X}^i,{\bf X}^j] \right)\,.
\ee
This Hamiltonian gives rise to the Matrix equations of motion
\[
\ddot{\bf X}^i + [[{\bf X}^i,{\bf X}^j],{\bf X}^j] = 0
\]
which must be supplemented with the Gauss constraint
\be
\label{MatrixGauss}
[\dot{\bf X}{}^i, {\bf X}^i] = 0 \,.
\ee

As a particularly simple example of this formalism, which will be
useful to us later, we consider the dynamics of an ``ellipsoidal''
membrane, in which the worldvolume sphere is mapped linearly into the
first three transverse coordinates.  First we introduce an equal-area
parameterization of the unit sphere, for example by coordinates
$\sigma^a = (x, \theta)$ with $-1 \le x \le 1$, $\theta \approx \theta
+ 2 \pi$.  The Cartesian coordinates on the sphere
\begin{eqnarray*}
x_1 & = &  x\\
x_2 & = &  \sqrt{1-x^2} \sin \theta\\
x_3 & = & \sqrt{1-x^2} \cos\theta
\end{eqnarray*}
obey the expected $SU(2)$ algebra: $\{x_A,x_B\} = \epsilon_{ABC} x_C$.
The ellipsoid is described by setting
\be
X_i(t,\sigma^a) = r_i(t) x_i(\sigma^a), \;\;\;\;\; i \in\{1, 2, 3\} \,.
\label{eq:ellipsoid}
\ee
By taking the velocities $\dot{r}_i$ to vanish at some initial time,
we are guaranteed to satisfy the constraint (\ref{MembraneGauss}).
The equations of motion for the radii are
\begin{eqnarray}
\ddot{r}_1 & = &  - \alpha r_1 (r_2^2 + r_3^2) \nonumber\\
\ddot{r}_2 & = &  - \alpha r_2 (r_1^2 + r_3^2) \label{eq:ellipsoid-eom}\\
\ddot{r}_3 & = &  - \alpha r_3 (r_1^2 + r_2^2) \nonumber
\end{eqnarray}
where $\alpha = 4/N^2$.  As expected, an initially static membrane
will start to contract.  If the membrane is initially an isotropic sphere of
radius $r_0$, energy conservation gives
\[
\dot{r}^2(t) + \alpha r^4 = \alpha r_0^4 \,.
\]
Classically, after a time
\[
t =  {1 \over r_0} \, {N \Gamma(1/4)^2 \over \sqrt{128 \pi}}\,,
\]
a spherical membrane contracts to a point and begins to expand again
with the opposite orientation.  This solution was originally described
by Collins and Tucker \cite{ct}.

It is also straightforward to solve the constraint equations
(\ref{constraints}) to find the behavior of the general ellipsoidal membrane in
$X^-$.  One finds that
\[
X^-(t,\sigma^a) = {p^- \over p^+} t + \half \sum_A r_A \dot{r}_A
\left(x_A^2(\sigma^a) - {1 \over 3}\right)\,.
\]
Note that an initially spherical membrane will remain spherical as it
contracts.  But in general an ellipsoidal membrane will begin to
fluctuate in $X^-$.

For linearly embedded membranes, the correspondence with matrix
theory is both simple and exact, even at finite $N$.  The matrices associated
with the ellipsoidal membrane
\be
{\bf X}_i(t) = {2 \over N} \, r_i(t) J_i, \;\;\;\;\; i \in\{1, 2, 3\}
\label{eq:MatrixEllipsoid}
\ee
obey commutation relations which are proportional to the Poisson brackets
obeyed by the classical membrane coordinates.  So the equations of motion
will agree exactly, even at finite $N$.

This exact agreement is a special property of ellipsoidal membranes,
or more generally of membranes with linear transverse embeddings.
Generally speaking, the Poisson bracket of two polynomials of degree
$m$ and $n$ is a polynomial of degree $m + n - 1$.  Thus, if $X_i$
contains terms of degree larger than 1, over time terms of arbitrarily
large degree will be excited, which clearly makes an exact
correspondence between membranes and finite-$N$ Matrix theory
impossible.  This will be discussed further in section \ref{sec:eom}.

\subsection{Membrane geometry at finite $N$}
\label{sec:geometry}

We now make a few observations which may help to clarify the
geometrical interpretation of finite-$N$ spherical matrix membranes.
For simplicity consider a spherical membrane embedded isotropically in
the first three transverse space coordinates.
\[
X_i(t,\sigma^a) = r(t) x_i(\sigma^a) \;\;\;\;\; i \in \{1, 2, 3\}\,.
\]
The corresponding matrices are
\[
{\bf X}_i(t) = {2 \over N} r(t) J_i, \;\;\;\;\; i \in \{1, 2, 3\}\,.
\]
According to the usual Matrix theory
description \cite{BFSS}, the matrices ${\bf X}_i$
describe a set of $N$ 0-branes (supergravitons with unit momentum in
the longitudinal direction) which have been bound together with
off-diagonal strings.  If this matrix configuration is truly to
describe a macroscopic spherical membrane, we would expect that the
$N$ 0-branes should be uniformly distributed on the surface of the
sphere of radius $r$.  A first check on this is to calculate the sum of
the squares of the matrices.
\[
{\bf X}_1^2 + {\bf X}_2^2 + {\bf X}_3^2 = r^2 \, \left(1 - \frac{1}{N^2}
\right)\,\identity\,.
\]
This indicates that for large $N$ the 0-branes are in a noncommutative
sense constrained to lie on the surface of a sphere of radius $r$.
The radius of the matrix sphere, defined in this way, receives
finite-$N$ corrections from its classical ($N \rightarrow \infty$)
value.  This is a general feature: different definitions of the
geometry will not agree at order $1/N$, reflecting the intrinsic
fuzziness of a finite-$N$ matrix sphere.

A related geometrical check can be performed by diagonalizing any one of
the matrices ${\bf X}_i$.  If the matrix configuration is indeed to
describe a sphere, each matrix should have a distribution of
eigenvalues which matches the density of a uniform sphere
projected onto a coordinate axis.  This gives a density which is
uniformly distributed over the interval $[-r, r]$.  Indeed, the
eigenvalues of each of the $SU(2)$ generators are given by
\[
\frac{N - 1}{2},\enskip \frac{N - 3}{2},\ldots, \frac{3 - N}{2},
\enskip \frac{1 - N}{2} 
\]
and thus the spectrum of eigenvalues of each matrix ${\bf X}_i$ is given by
\[
r(1-1/N), \enskip r (1-3/N), \ldots, r (-1+ 3/N), \enskip r (-1 + 1/N).
\]
If we associate a 0-brane with each eigenvalue, and we assume that
each 0-brane corresponds to an equal quantum of area on the surface of
the sphere, we see that the area associated with these 0-branes
precisely covers the interval $[-r, r]$ with a uniform distribution.
In contrast to the above, this is consistent with the classical
expectation, with no $1/N$ corrections.

As another example of how the spherical membrane geometry can be
understood from the matrices, we ask whether the 0-branes
corresponding to a small patch of the sphere can be studied in
isolation.  For large $r$ and $N$, we expect that we should be
able to isolate a region which looks locally like an infinite planar
membrane, by focusing on the part of the matrices in which one of the
coordinates is close to its maximum value (for example, $X_3 \approx
r$).

Indeed, this is the case.  If we work in an eigenbasis of $J_3$ with
eigenvalues decreasing along the diagonal, then the matrices ${\bf
X}_1$, ${\bf X}_2$ satisfy
\[
[\bfx_1, \bfx_2] =  {2 i r^2 \over N} \left(\begin{array}{ccccc}
1  -\frac{1}{N}  & 0 & 0 & \ddots &\\
0 &1  -\frac{3}{N}  & 0 & 0 & \ddots\\
0 & 0 &1  -\frac{5}{N}  & 0 & \ddots\\
 \ddots & 0 & 0 &1-\frac{7}{N}  & \ddots\\
& & \ddots & \ddots & \ddots
\end{array}\right)
\]
Restricting attention to the upper left-hand corner, up to terms of
order $1/N$ this reduces to $[{\bf X}_1, {\bf X}_2] = {2 i r^2
\over N} \identity$.  This is precisely the type of relation
introduced in BFSS \cite{BFSS} to describe an infinite planar
membrane stretched in the $X^1 X^2$ plane.  Thus, in a sense this
subset of the 0-branes describes a geometrical ``piece'' of the full
spherical membrane.

\subsection{Equations of motion at finite $N$}
\label{sec:eom}

As we have seen, finite-$N$ Matrix theory has states which can be
identified geometrically with membranes of spherical topology.  To
complete the identification we must further establish that these
states in Matrix theory evolve according to the equations of motion of
the corresponding membranes.

The membrane and Matrix Hamiltonians (\ref{MembraneHamiltonian}),
(\ref{MatrixHamiltonian}) are related by the Poisson bracket -- matrix
commutator correspondence
\[
\{\cdot,\cdot\} \leftrightarrow {-i N \over 2} [\cdot,\cdot]\,.
\]
Note that $2/N$ plays the role of $\hbar$ in the usual
classical-quantum correspondence.  Thus in the large-$N$
limit one expects the two equations of motion to agree.

It is important, however, to analyze the corrections which result from
finite $N$.  Not surprisingly finite-$N$ Matrix theory places a
restriction on the number of modes which can be excited on the
membrane.  Suppose $Z_1(x_A)$ and $Z_2(x_A)$ are two polynomials of
degree $k_1$ and $k_2$, respectively.  Then their Poisson bracket $Z_3
= \{Z_1,Z_2\}$ is a polynomial of degree $k_3 = k_1 + k_2 - 1$.  The
associated matrices obey
\[
-{i N \over 2} [{\bf Z}_1,{\bf Z}_2] = {\bf Z}_3 \left(1 + {\cal O}\left({k_1
k_2 / N}\right)\right)\,.
\]
So, in order for the matrix and membrane equations of motion to agree,
the modes which are excited on the membrane must satisfy
$k \ll k_{\rm max} \sim \sqrt{N}$.  Note that this is a stronger condition
than the requirement $k < N$ which was necessary for the membrane -- matrix
correspondence (\ref{MMcorrespond}) to generate linearly independent matrices.

This restriction, even if satisfied at some initial time, will not in
general persist.  The membrane equations of motion are non-linear, so
higher and higher modes on the membrane will become excited.  This is
just a consequence of the fact that for $k_1,\,k_2 > 1$ taking the
Poisson bracket generates polynomials of increasing degree.  But an
initially large, smooth membrane should be well described by matrix
theory for a long time before the non-linearities become important.
Note that the ellipsoidal membranes discussed above are an exception
to this general rule: the classical matrix and membrane equations
agree exactly for arbitrary $N$ and for all time.

So we see that two conditions must be satisfied for a state in matrix
theory to behave like the corresponding semiclassical membrane.
First, the size of the membrane must be much larger than the Planck
scale, so that the semiclassical approximation to the Matrix equations
of motion is valid.  Second, the modes which are excited on the
membrane must obey the restriction $k < \sqrt{N}$.  This latter
restriction is analogous to the condition $S < N$ which was found to
be necessary to have an accurate finite-$N$ description of a matrix
black hole
\cite{bfks,Klebanov-Susskind,Horowitz-Martinec,Bigatti-Susskind}.  The
restriction can be understood heuristically in the present context by
noting that finite-$N$ Matrix theory describes the dynamics of $N$
partons, which matches the $N$ degrees of freedom of a membrane that
only has modes up to $k \sim \sqrt{N}$ excited.

\section{Matrix theory potential}

As we have seen, Matrix theory correctly reproduces the classical
dynamical equations which govern an M-theory membrane.  Of course the
spacetime interpretation requires that the membrane be surrounded by a
gravitational field.  We now wish to show that Matrix theory correctly
reproduces this field, at least at large distances.  To do this, we
will compute the force exerted by a membrane on a D0-brane or  on
another membrane located far away.  Closely related calculations have
been performed by many authors; see for example
\cite{BFSS,DKPS,Aharony-Berkooz,Lifschytz-Mathur,Lifschytz-46,bc,ChepTseyI,MaldacenaI,Vakkuri-Kraus,Gopakumar-Ramgoolam,ChepelevTseytlin,MaldacenaII,Esko-Per2}.

It is natural at this point to make the discussion somewhat more general.
Consider any state in Matrix theory which, from the spacetime point
of view, is localized within a finite region.  Then we expect a
long-range gravitational field to be present, with leading
long-distance falloff governed by a conserved quantity: the total mass
of the system.  This should also hold true in Matrix theory.  That
is, the leading long-distance potential seen by any probe should be
governed by a conserved quantity, which can be identified with the
mass of the system.


With this as motivation, we introduce a background
\be
\label{background}
\langle {\bf X}_i (t) \rangle = \left[
\ba{cc}
\bfy_i(t) & 0 \\
0 & \tilde{\bfy}_i(t)
\ea
\right]
\ee
into the Matrix Yang-Mills theory.  This background describes two
systems.  $\bfy_i(t)$ $i=1,\ldots,9$ is an $N \times N$ matrix
representing the first system, while $\tilde{\bfy}_i(t)$ is an $\tilde{N}
\times \tilde{N}$ matrix representing the second system.  In principle
this background should be taken to satisfy the Matrix equations of
motion, although for our purposes explicit solutions will not turn out
to be necessary.  We assume that the systems are localized, in that
the spreads in the eigenvalues of the matrices $\bfy_i$, $\tilde{\bfy}_i$ are
all much smaller than the magnitude of the vector between their
transverse center of mass positions, which we denote by $b_i$.

To compute the interaction potential we must integrate out the off-block-diagonal
elements of the matrices.  So we set
\[
A_0 = \left[ \ba{cc}
0 & {a} \\
{a}^{\dag} & 0
\ea \right]
\hskip 1.0 cm
{\bf X}_i = \langle {\bf X}_i \rangle + \left[ \ba{cc}
0 & {x_i} \\
{x_i}^{\dag} & 0
\ea \right]
\hskip 1.0 cm
\Psi_a = \left[ \ba{cc}
0 & {\psi_a} \\
{\psi_a}^{\dag} & 0
\ea \right]
\]
where the fluctuations $a$, $x_i$, $\psi_a$ are $N \times \tilde{N}$
matrices.  The Matrix action reads\footnote{ Conventions: All fields
are Hermitian.  $i,j = 1,\ldots,9$ are $SO(9)$ vector indices with
metric $\delta_{ij}$, while $a,b=1,\ldots,16$ are $SO(9)$ spinor
indices.  The $SO(9)$ Dirac matrices $\gamma^i$ are real, symmetric,
and obey $\left\lbrace \gamma^i, \gamma^j \right\rbrace = 2
\delta^{ij}$, $\gamma^1 \cdots \gamma^9 = + \identity$.  We define
$\gamma^{ij} = \half [\gamma^i,\gamma^j]$.}
\beas
S &=& {1 \over \gym} \int dt \tr \biggl\lbrace D_t {\bf X}_i D_t {\bf X}_i
+ \half [{\bf X}_i,{\bf X}_j] [{\bf X}_i,{\bf X}_j]
- \left(D_\mu^{\rm cl} A^{\mu \, {\rm qu}}\right)^2 \\
&& \qquad \qquad \qquad + i \Psi_a D_t \Psi_a - \Psi_a \gamma^i_{ab}
[{\bf X}_i,\Psi_b] \biggr\rbrace
\eeas
where we have added a covariant background-field gauge fixing term
\[
D_\mu^{\rm cl} A^{\mu \, {\rm qu}} \equiv - \partial_t A_0 + i
[\langle {\bf X}_i \rangle, {\bf X}_i]\,.
\]
The corresponding ghosts will be taken into account below.  The Yang-Mills coupling
is fixed by matching to the Hamiltonian (\ref{MatrixHamiltonian}): $\gym = 2 R$.

We are interested in the interaction potential between the system and
the probe at a given instant of time.  This potential is simply the
ground state energy of the off-block-diagonal fields ${a}$,
${x}_i$, ${\psi}_a$.
The leading
large-distance potential arises at one loop in Matrix theory.  Thus, we
expand the Matrix action to quadratic order in the
off-block-diagonal fields.  We end up with a system of harmonic
oscillators, whose frequencies are determined by the background fields
$\bfy_i(t)$ and $\tilde{\bfy}_i(t)$ (along with their time derivatives $\partial_t
\bfy_i$, $\partial_t \tilde{\bfy}_i$).  The Yang-Mills plus gauge-fixing terms give
rise to $10 N\tilde{N}$ complex bosonic oscillators with $({\rm
frequency})^2$ matrix
\beas
\left(\Omega_b\right)^2 &=& M_{0\,b} + M_{1\,b} \\
\noalign{\vskip 0.2 cm}
M_{0\,b} &=& \sum_i K_i^2 \otimes \identity_{10 \times 10} \\
M_{1\,b} &=& \left[ \ba{cc}
0 & - 2  \partial_t K_j \\
2 \partial_t K_i & 2 [K_i,K_j] \ea \right]
\eeas
where
\[
K_i \equiv \bfy_i \otimes \identity_{\tilde{N} \times \tilde{N}} 
-  \identity_{N \times N} \otimes \tilde{\bfy}_i^T\,.
\]
Similarly, the
fermions give rise to $16 N\tilde{N}$ complex fermionic oscillators with $({\rm
frequency})^2$ matrix
\beas
\left(\Omega_f\right)^2 &=& M_{0\,f} + M_{1\,f} \\
\noalign{\vskip 0.2 cm}
M_{0\,f} &=& \sum_i K_i^2 \otimes \identity_{16 \times 16} \\
M_{1\,f} &=& i \partial_t K_i \otimes \gamma^i
+ \half [K_i,K_j] \otimes \gamma^{ij}
\eeas
Finally we have (two identical sets of) $N\tilde{N}$ complex scalar ghost
oscillators with $({\rm frequency})^2$ matrix
\[
\left(\Omega_g\right)^2 = \sum_i K_i^2\,.
\]

We are only interested in the leading large-distance potential.  Then
these oscillators have very large frequencies, on the order of the
separation distance $b$, and we can treat the background as if it is
quasi-static by comparison.  That is, we can pretend that the
background fields $\bfy_i$, $\tilde{\bfy}_i$ (as well as their time
derivatives!) are constant matrices.  Then the ground state energy, or
equivalently the interaction potential, is given by the usual
oscillator formula
\be
\label{potential}
V_{\rm matrix} = \tr \left(\Omega_b\right) - \half \tr \left(\Omega_f\right)
- 2 \tr \left(\Omega_g\right)\,.
\ee

We are interested in the leading behavior of the potential
as $b \rightarrow \infty$.  Thus we can
treat $M_1$ as a perturbation, since its eigenvalues stay of order one
in this limit, while the eigenvalues of $M_0$ are of order $b^2$.

To make the expansion, it is convenient to first introduce an integral
representation for the square root of a matrix, then use the standard
Dyson perturbation series.
\beas
\lefteqn{\tr \sqrt{M_0 + M_1} } \\
& = & - {1 \over 2 \sqrt{\pi}} \tr
\int_0^\infty {d \tau \over \tau^{3/2}} \, e^{- \tau (M_0 + M_1)} \\
& = & - {1 \over 2 \sqrt{\pi}} \trg \biggl\lbrace
\int_0^\infty {d \tau_1 \over \tau_1^{3/2}} \, e^{-\tau_1 M_0} \trl \Bigl(\identity\Bigr)
- \int_0^\infty\int_0^\infty {d \tau_1 d \tau_2 \over \left(\tau_1+\tau_2\right)^{3/2}} \,
e^{-(\tau_1 + \tau_2) M_0} \trl \Bigl(M_1(\tau_2)\Bigr) \\
& & \quad + \int_0^\infty \int_0^\infty\int_0^\infty
{d \tau_1 d \tau_2 d \tau_3 \over \left(\tau_1 + \tau_2 + \tau_3
\right)^{3/2}} \, e^{-(\tau_1 + \tau_2 + \tau_3) M_0} \trl
\Bigl(M_1(\tau_2+\tau_3) M_1(\tau_3)\Bigr) \mp \cdots \biggr\rbrace\,.
\eeas
In the last line we have broken the trace up into $\tr = \trg \trl$,
where $\trg$ is an $N \tilde{N}$-dimensional trace on group indices, and $\trl$
is an appropriate trace on Lorentz indices.  Also we have defined the `interaction
picture' field
\[
M_1(\tau) \equiv e^{\tau M_0} M_1 e^{-\tau M_0}\,.
\]

It is now straightforward to expand the potential.  At zeroth order in
the perturbation we get the expected supersymmetric cancellation of
zero-point energy.
\[
\trl \Bigl(\identity_{10 \times 10}\Bigr) - \half \trl \Bigl(\identity_{16 \times 16}\Bigr)
- 2 = 0\,.
\]
Note that this cancellation arises from the trace over Lorentz
indices, and does not depend on the detailed form of the matrices
$K_i$.  We shall see that this feature persists at higher orders.
Also, note that if we were dealing with a configuration in which the
background fields $\langle {\bf X}_i \rangle$ commuted and were
constant, then this lowest-order result would not be corrected: the
perturbation $M_1$ vanishes.  This is the expected BPS result that
there is no potential between a system of stationary D0-branes.

At first order in the perturbation, the bose and fermi contributions separately
vanish:
\[
{\rm Tr_L} \Bigl(M_{1\,b}(\tau)\Bigr) = 0 \hskip 1.0 cm \trl \Bigl(
M_{1\,f}(\tau)\Bigr) = 0\,.
\]
At second order there is a cancellation:
\beas
\lefteqn{ {\rm Tr_L} \Bigl(M_{1\,b}(\tau_1) M_{1\,b}(\tau_2)\Bigr) 
= \half {\rm Tr_L} \Bigl(M_{1\,f}(\tau_1) M_{1\,f}(\tau_2)\Bigr)} \\
&=&- 8 \, \partial_t \,\left.K_i\right\vert_{\tau_1} \partial_t
\,\left.K_i\right\vert_{\tau_2} - 4
\left.[K_i,K_j]\right\vert_{\tau_1}
\left.[K_i,K_j]\right\vert_{\tau_2} \eeas where for example we have
defined $\left.K_i\right\vert_\tau = e^{\tau M_0} K_i e^{- \tau M_0}$.
At third order the cancellation is less trivial: \beas \lefteqn{{\rm
Tr_L} \Bigl(M_{1\,b}(\tau_1) M_{1\,b}(\tau_2) M_{1\,b}(\tau_3)\Bigr) =
\half {\rm Tr_L} \Bigl(M_{1\,f}(\tau_1) M_{1\,f}(\tau_2)
M_{1\,f}(\tau_3)\Bigr)} \\
&=& -8 \left.[K_i,K_j]\right\vert_{\tau_1} \partial_t \!\left.K_j\right
\vert_{\tau_2} \partial_t \!\left.K_i\right\vert_{\tau_3}
-8 \, \partial_t \!\left.K_i\right\vert_{\tau_1} \left.[K_i,K_j]\right\vert_{\tau_2}
\partial_t \!\left.K_j\right\vert_{\tau_3}  \\
& & - 8 \, \partial_t \!\left.K_i\right\vert_{\tau_1}
\partial_t \!\left.K_j\right\vert_{\tau_2} \left.[K_j,K_i]\right\vert_{\tau_3}
+8 \left.[K_i,K_j]\right\vert_{\tau_1} \left.[K_j,K_k]\right\vert_{\tau_2}
\left.[K_k,K_i]\right\vert_{\tau_3}\,.
\eeas

At fourth order we find a non-zero result.  At large $b$ the integral
is dominated by the small-$\tau$ behavior of the integrand, so we can
expand and find the leading term
\bea
\label{MatrixPotential}
V_{\rm matrix} &=&
 - {1 \over 2 \sqrt{\pi}} \trg \int_0^\infty {d \tau_1 \cdots d\tau_5 \over
\left(\tau_1 + \cdots + \tau_5\right)^{3/2}} \, e^{- (\tau_1 + \cdots + \tau_5)
b^2} \left(
\trl \Bigl(M_{1\,b}^4\Bigr) - \half \trl \Bigl(M_{1\,f}^4\Bigr)\right) \nonumber \\
&=& - {5 \over 128 \, b^7} \, W
\eea
where the ``gravitational coupling'' $W$ is given by
\bea
\label{W}
W & \equiv & \trg \left(\trl \Bigl(M_{1\,b}^4\Bigr) - \half \trl
\Bigl(M_{1\,f}^4\Bigr)\right) \label{eq:w}\\
&=& {\rm Tr}_G \biggl\lbrace
   8 F^\mu{}_\nu F^\nu{}_\lambda F^\lambda{}_\sigma F^\sigma{}_\mu
+ 16 F_{\mu \nu} F^{\mu\lambda} F^{\nu\sigma} F_{\lambda \sigma}
-  4 F_{\mu\nu} F^{\mu \nu} F_{\lambda \sigma} F^{\lambda \sigma}
-  2 F_{\mu\nu} F_{\lambda \sigma} F^{\mu\nu} F^{\lambda \sigma}
\biggr\rbrace \,.\nonumber
\eea
We have introduced the notation $F_{0i} = - F_{i0} = \partial_t K_i$
and $F_{ij} = i [K_i, K_j]$.  This general result is discussed in
\cite{ChepelevTseytlin1,ChepelevTseytlin}.  A similar formula was
discussed in \cite{MaldacenaI,MaldacenaII}, but with the assumption
that commutators of field strengths could be dropped.

\section{Comparison to supergravity}

We now compare the general Matrix expression (\ref{MatrixPotential})
to the naive supergravity potential, that is, to the potential arising
from tree-level graviton exchange in discrete light-front
supergravity.

\subsection{Matrix and gravitational potentials}
\label{sec:potentials}

In general, the effective action from one-graviton exchange in a flat
background metric $\eta_{\mu\nu} = (- + \cdots +)$ is
\[
S_{eff} =  - {1 \over  8}\int d^dx d^dy \, T_{\mu\nu}(x)
D^{\mu\nu\lambda\sigma}(x-y) T_{\lambda\sigma}(y)
\]
where the (harmonic gauge) graviton propagator in $d$ spacetime dimensions is
\[
D^{\mu\nu\lambda\sigma}(x-y) = 16 \pi G \left(\eta^{\mu\lambda}
\eta^{\nu\sigma} + \eta^{\mu \sigma} \eta^{\nu \lambda} - {2 \over d-2}
\eta^{\mu \nu} \eta^{\lambda \sigma} \right) \int {d^dk \over (2 \pi)^d}
{e^{i k \cdot (x-y)} \over - k^2}\,.
\]
We should use a Green's function which is periodic in $X^-$ but this
turns out not to affect the result.  The stress tensor for an
object with momentum $p^\mu$ located at light-front coordinates
$(z^-,z^i)$ is
\begin{equation}
T^{\mu\nu}(x^+,x^-,x^i) = {p^\mu p^\nu \over p^+} \, \delta(x^--z^-)
\delta(x^i-z^i)\,.
\label{eq:stress-energy}
\end{equation}
With two objects the total stress tensor is a sum $T^{\mu \nu}(x) +
\tilde{T}^{\mu \nu}(x)$.  In accord with our quasi-static
approximation in Matrix theory, we treat the positions $z$,
$\tilde{z}$ as if they are constants.  To get the potential from the
effective action we change the overall sign and strip off $\int dx^+$.
Because
no longitudinal momentum is exchanged, we must 
average ${1 \over 2 \pi R} \int dx^-$.  This gives
\be
\label{GravityPotential}
V_{\rm gravity} = - {15 \over 4} \, {R^4 \over N \tilde{N} b^7}
\left[\left(p \cdot \tilde{p}\right)^2 - {1 \over 9} p^2 \tilde{p}^2\right]
\ee
where we have specialized to eleven dimensions.  Our normalizations are such
that $16 \pi G = (2 \pi)^8 l_P^9$ \cite{dealwis-normalize}, and we use units in which
$2 \pi l_P^3 = R$.

We can compare this to the predictions of Matrix theory in two
cases.  First, consider the case where the second object is a massless particle with
\[
\tilde{p}^+ = \tilde{N}/R \hskip 1.0 cm \tilde{p}^- = 0 \hskip 1.0 cm
\tilde{p}^i = 0 \,.
\]
Then the gravitational potential simplifies to
\be
\label{massless-gravity}
V_{\rm gravity} = - {15 \over 4} \, {R^2 \tilde{N} \over b^7 N} \, E^2
\ee
where $E$ is the light-front energy of the first object, given by the
Matrix theory Hamiltonian
\be
\label{MatrixHamiltonian2}
E = {1 \over R} \tr \left( \half \dot{{\bf \bfy}}^i \dot{{\bf \bfy}}^i - {1 \over 4}
[{\bf \bfy}^i,{\bf \bfy}^j] [{\bf \bfy}^i,{\bf \bfy}^j] \right)\,.
\ee
The Matrix theory and supergravity predictions for the potential agree when
\begin{equation}
W = 96 \frac{R^2}{N} E^2\,.
\label{eq:compare-1a}
\end{equation}
Here $W$ is to be constructed just from the matrix variables $\bfy_i$
describing the first object ({\em i.e.} $F_{0i} =
\dot{\bfy_i}$ and $F_{ij} = i[\bfy_i, \bfy_j]$ in (\ref{eq:w})).

The second case is where both objects are massive.  We take both of
the objects to be at transverse rest.  Then the Matrix and
supergravity potentials agree when
\begin{equation}
W = 96 \frac{R^2}{N \tilde{N}} \left[(\tilde{N} E)^2
 + (N\tilde{E})^2 + \frac{14}{9} (\tilde{N} E)  (N\tilde{E}) \right]
\label{eq:compare-2a}
\end{equation}
where $W$ is constructed from $K_i =
\bfy_i \otimes\identity_{\tilde{N} \times \tilde{N}} - \identity_{N
\times N} \otimes \tilde{\bfy}_i^T$, $\bfy_i$ and $\tilde{\bfy}_i$ are
matrices describing the two objects, and the Matrix energies $E,
\tilde{E}$ are given by (\ref{MatrixHamiltonian2}) evaluated on $\bfy_i$,
$\tilde{\bfy}_i$ respectively.

We emphasize that we are comparing the Matrix theory potential
(\ref{MatrixPotential}) to the gravitational potential
(\ref{GravityPotential}) which we {\em evaluate using the Matrix expressions
for $p$ and $\tilde{p}$.}  This means we never have to
specify a precise correspondence between a state in Matrix theory and
a state in supergravity.  This is fortunate because, as we discussed
in section \ref{sec:geometry}, these identifications are afflicted with $1/N$
ambiguities.  Instead, we can test whether the expected relationship
(\ref{GravityPotential}) holds internally within Matrix theory for any $N$.

An important subtlety arises in verifying the relations
(\ref{eq:compare-1a}) and (\ref{eq:compare-2a}) which express the
correspondence between the Matrix and supergravity potentials.  For
many membrane configurations with nontrivial time dependence, the
Matrix theory potential is time-dependent.  Generally, this
time-dependence arises from the outgoing gravitational radiation
produced by the fluctuating membrane source(s).  In order to compare
with the supergravity calculation, where we assumed a stress-energy
tensor of the form (\ref{eq:stress-energy}), we must neglect
gravitational radiation.  This can be done in Matrix theory by simply
time-averaging the expression for $W$ in (\ref{W}).  In the remainder
of this paper we will time-average all Matrix potentials as needed in
order to remove radiation effects.  A more complete discussion of
gravitational radiation in Matrix theory will appear in
\cite{Dan-Wati2}.

To summarize the discussion, in order to check that a single object
described by matrices $\bfy_i$ has the proper interaction with a
graviton it suffices to verify the matrix relation
\begin{equation}
\langle W \rangle = 96 \frac{R^2}{N} E^2\,.
\label{eq:compare-1}
\end{equation}
where $\langle \cdot \rangle$ denotes a time average.  Similarly, to verify
that two massive objects described by matrices $\bfy_i$,
$\tilde{\bfy}_i$ have the proper long-range potential it suffices to
verify the relation 
\begin{equation}
\langle W\rangle = 96 \frac{R^2}{N \tilde{N}} \left[(\tilde{N} E)^2 +
 (N\tilde{E})^2 + \frac{14}{9} (\tilde{N} E) (N\tilde{E}) \right].
\label{eq:compare-2}
\end{equation}

The rest of this paper is devoted to checking these relations.  In
section \ref{sec:finite-N} we consider a simple example: a finite $N$
sphere interacting with a graviton.  We will find that the leading
large-$N$ potentials agree, but that there is a discrepancy at
subleading orders in $N$.  Then, in section \ref{sec:theorem-1}, we
prove a general result that (\ref{eq:compare-1}) always holds in a
formal large-$N$ limit.  In section \ref{sec:theorem-2} we show that,
even if $N$ is finite, any two objects which separately satisfy
(\ref{eq:compare-1}) will jointly satisfy (\ref{eq:compare-2}), as
long as the internal dynamics of the two objects are not correlated
over time.

\subsection{Sphere-graviton interaction}
\label{sec:finite-N}

The first thing we will examine is the long-range potential between a
finite-$N$ spherical membrane and a graviton.  As in section
\ref{sec:description} we describe the sphere by matrices
\[
\bfy_i(t) = \frac{2r(t)}{N}  J_i, \;\;\;\;\; i \in \{1, 2, 3\}.
\]
The corresponding background field strengths are
\[
F_{0i}=\frac{2\dot{r}}{N}  J_i, \;\;\;\;\;
F_{ij} = -\frac{4 r^2}{N^2}  \epsilon_{ijk} J_k.
\]
The Matrix energy of the sphere is given by (\ref{MatrixHamiltonian2}).
\beas
E &=& {1 \over R} \left(\frac{2 \dot{r}^2}{N^2} + \frac{8r^4}{N^4}\right) \;
{\rm Tr} \; \bigl(J_i J_i\bigr) \\
&=& {8 r_0^4 \over R N^3} \, c_2\,.
\eeas
Here $c_2 = \frac{N^2 -1}{4}$ is the quadratic Casimir, and $r_0$ is the initial
radius of the sphere: $\dot{r}^2 + {4 \over N^2} r^4 = {4 \over N^2} r_0^4$.
Thus from (\ref{massless-gravity}) we expect the gravitational potential
\[
V_{\rm gravity} = - 240 {r_0^8 \over N^7 b^7} c_2^2\,.
\]

On the other hand we can compute the one-loop Matrix potential.  Substituting
the field strengths into (\ref{eq:w}) we find that the coupling $W$ is given by
\beas
W &=& {2^{12} r_0^8 \over N^8} \left[\tr \left(J_i J_i J_j J_j\right)
+ \half \tr \left(J_i J_j J_i J_j\right)\right] \\
&=& {2^{12} r_0^8 \over N^8} \left({3 \over 2} N c_2^2 - \half N c_2\right)
\eeas
and thus from (\ref{MatrixPotential}) that the Matrix potential is
\[
V_{\rm matrix} = - 240 {r_0^8 \over N^7 b^7} \left(c_2^2 - {1 \over 3} c_2\right) \,.
\]
Since $c_2$ is order $N^2$, we see that the large-$N$ behavior of the
two potentials is identical.  This supports the BFSS conjecture that
the large-$N$ limit of Matrix quantum mechanics reproduces
uncompactified M-theory and its low-energy limit, 11-dimensional
supergravity.

Following the discrete light-front conjecture
\cite{Susskind-DLCQ}, however, we might also expect agreement at finite $N$.
This is evidently not the case.  The ordering of the matrices in the
expression for the Matrix potential (\ref{MatrixPotential}) gives rise
to a term, involving the commutator of two field strengths, which is
subleading in $N$.  But this term is not present in the naive
supergravity potential (\ref{GravityPotential}).

It seems unlikely that our lowest-order results will be corrected in a
way that resolves this discrepancy.  For example, there are
higher-derivative corrections to 11-dimensional supergravity, which
modify the graviton propagator.  But such modifications produce
effects which fall off more rapidly than $1/b^7$.  Likewise, there are
loop corrections in the Matrix gauge theory.  But such corrections
will bring in additional powers of the dimensionless quantities $\gym
/ b^3$ and $\gym / r^3$.

Very convincing arguments have been advanced that finite-$N$ Matrix
theory provides a description of DLCQ M-theory
\cite{Seiberg-DLCQ,dealwis-DLCQ}.  So a plausible resolution seems to
be that the DLCQ of M-theory is simply not described at low energies
by the DLCQ of 11-dimensional supergravity \cite{banks-review,hp}.  In
particular, there is no reason to believe that DLCQ M-theory should
have a low energy effective Lagrangian which is Lorentz invariant, or
even local.

\subsection{Large $N$ interactions with a graviton}
\label{sec:theorem-1}

We will now prove a general theorem stating that an arbitrary
finite-size classical membrane configuration in Matrix theory has the
correct long-distance interactions with a graviton at large $N$.  We
proceed as follows:  Our goal is to establish that
(\ref{eq:compare-1}) holds at large $N$.\footnote{This procedure
amounts to taking the limit of large separation first, then the limit
of large $N$.  It is a subtle question whether the opposite order of
limits, which is what the IMF proposal of BFSS \cite{BFSS} calls for,
will give the same result.  We thank Tom Banks for emphasizing this
point to us.}  We do this formally, in two steps.  First, we invert
the correspondence between membranes and matrices
(\ref{MMcorrespond}).
\[
\{\cdot,\cdot\} \leftrightarrow {-i N \over 2} [\cdot,\cdot] \hskip 1.0 cm
{N \over 4 \pi} \int d^2 \sigma \leftrightarrow \tr
\]
Then we make use of the classical membrane equations of motion
discussed in section \ref{sec:description} to simplify the resulting
expressions.  Of course these formal manipulations are only valid on a
limited class of matrices, namely those which converge to a smooth
membrane in the large-$N$ limit.  We leave it to future work to make
the proof more general.

Translating back into membrane language, in the expression for $W$
(\ref{W}) we replace the matrices $\bfy_i(t)$ with functions
$Y_i(t,\sigma^a)$, and identify
\[
F_{0i} = \dot{Y}_i \hskip 1.0 cm F_{ij} = - {2 \over N} \{Y_i,Y_j\}\,.
\]
The resulting expression simplifies, with the use of the identity
\[
\{Y_i,Y_j\} \{Y_j,Y_k\} = - \gamma \gamma^{ab} \partial_a Y_i \partial_b Y_k \, ,
\]
to
\beas
W &=&\;  {N \over 4 \pi} \int d^2 \sigma \left( 24
\; \dot{Y}_i \dot{Y}_i \dot{Y}_j \dot{Y}_j +
{192 \over N^2} \gamma \dot{Y}_i \dot{Y}_i + {384 \over N^4} \gamma^2
- {384 \over N^2} \gamma \gamma^{ab} \dot{Y}_i \bigl(\partial_a Y_i\bigr)
\dot{Y}_j \bigl(\partial_b Y_j\bigr)  \right)\\
&=& {N \over 4 \pi} \int d^2 \sigma \, \left(
 96 \left(\dot{Y}^-\right)^2 - {384 \over
N^2} \gamma \gamma^{ab} \partial_a Y^- \partial_b Y^- \right)
\eeas
where in the second line we made use of the constraint equations
for $Y^-$.  The constraints on $Y^-$ also imply that
\[
Y^-(t,\sigma^a) = {R \over N} E t + \xi(t,\sigma^a)
\]
where $E$ is the classical light-front energy of the object, and the
fluctuation satisfies $\int d^2 \sigma \; \xi = 0$.  In terms of $\xi$ we
have
\[
W =  {N \over 4 \pi} \int d^2 \sigma \, \left(
96 {R^2 \over N^2} E^2 + 96 {\partial
\over \partial t} \left(\xi \dot{\xi}\right) - 96 \xi \left[
\ddot{Y}^- - {4 \over N^2} \partial_a \left( \gamma \gamma^{ab} \partial_b Y^-
\right) \right] \right) \,.
\]
The final term in $W$
vanishes by the constraint equations for $Y^-$.  
As discussed in Section \ref{sec:potentials}, time-dependent terms in
$W$ correspond to gravitational radiation.  We wish to drop such terms
in checking (\ref{eq:compare-1}) to compare with supergravity.  We
remove radiation effects by integrating over time.
The fluctuation $\xi$ is of bounded variation, so the
term $\partial_t\left(\xi\dot{\xi}\right)$ will not contribute when
integrated over time, and can be dropped.  Thus we have shown
that after time-averaging
\[
\langle W \rangle = 96 {R^2 \over N} E^2
\]
which establishes (\ref{eq:compare-1}).

\subsection{Interactions between massive objects}
\label{sec:theorem-2}

We now prove the general result that any two objects
which separately have the correct long-range potentials with a
graviton in Matrix theory, also have the correct long-range potential
between themselves.  Mathematically, this amounts to proving that if
two sets of matrices $\bfy_i, \tilde{\bfy}_i$ each independently
satisfy (\ref{eq:compare-1}) then the two sets of matrices also
satisfy (\ref{eq:compare-2}).  In order to prove this theorem, we will
time-average the Matrix expression separately on each of the two sets
of matrix variables.  In the Matrix
theory picture this corresponds to the assumption that the internal
degrees of freedom describing the two objects are not correlated in
time so that
\[
\langle f (\bfy) g (\tilde{\bfy}) \rangle =
\langle f (\bfy) \rangle \cdot \langle g (\tilde{\bfy}) \rangle
\]
where the brackets denote time-averaging.  In general, this step is
justified unless there are correlations between the dynamics of the
two Matrix objects.  Generically, there will be no such correlations.
The behavior of the Matrix theory variables describing each object are
classically chaotic \cite{amrv}.  It is possible to construct
configurations with correlated dynamics, such as a pair of identical
spheres.  This corresponds physically to a pair of pulsating spherical
membranes whose periods of pulsation are equal.  Since the
gravitational radiation from each membrane will have the same period,
the interaction between each sphere and the radiation field of the
other will not  average to zero over time, so the gravitational
calculation is more complicated.

We now proceed with the proof in the case where the objects are
uncorrelated.  The Matrix gravitational coupling $W$
between the two objects can be broken into three terms, corresponding
to the three terms on the RHS of (\ref{eq:compare-2}).  The terms
which depend only on $\bfy_i$ or only on $\tilde{\bfy}_i$ are
correctly reproduced, as follows directly from (\ref{eq:compare-1}).
The only term which needs to be checked in detail is the cross term
containing matrices of both types.  Expanding out the terms in this
part of $W$ we find
\begin{eqnarray}
W_{\rm cross} \!\! & = & \!
{\rm Tr}\; \left\{
  48 \; \dot{\bfy}_i\dot{\bfy}_i \dot{\tilde{\bfy}}{}_j\dot{\tilde{\bfy}}{}_j  
+ 96 \; \dot{\bfy}_i\dot{\bfy}_j \dot{\tilde{\bfy}}{}_i\dot{\tilde{\bfy}}{}_j
+ 24 \; \dot{\bfy}_i\dot{\bfy}_i \tilde{F}{}_{jk}\tilde{F}{}_{jk}
+ 24 \; F_{jk}F_{jk} \dot{\tilde{\bfy}}{}_i\dot{\tilde{\bfy}}{}_i \right. \label{eq:cross-terms} \\ 
& & \qquad \left.
+ 96 \; \dot{\bfy}_i\dot{\bfy}_j \tilde{F}{}_{ik}\tilde{F}{}_{kj}
+ 96 \; F_{ik}F_{kj} \dot{\tilde{\bfy}}{}_i\dot{\tilde{\bfy}}{}_j
+ 96 \; F_{ij}F_{jk} \tilde{F}{}_{kl}\tilde{F}{}_{li}
- 12 \; F_{ij}F_{ij} \tilde{F}{}_{kl}\tilde{F}{}_{kl}\right\}
\nonumber
\end{eqnarray}
where we have abbreviated $F_{ij} = i[\bfy_i, \bfy_j]$.  In writing
(\ref{eq:cross-terms}) we have already made several simplifications.
A term proportional to $\dot{\bfy}_iF_{ij}$ has been dropped by the
Gauss constraint (\ref{MatrixGauss}):
\[
{\rm Tr}\; \left( \dot{\bfy}_i [\bfy_i, \bfy_j] \right) = {\rm Tr}\;
\left([\dot{\bfy}_i,\bfy_i] \bfy_j \right) = 0\,.
\]
A term proportional to $\dot{\bfy}_iF_{jk}$ has been dropped because
it is a total time derivative:
\[
\langle
{\rm Tr}\; \left( \dot{\bfy}_i F_{jk} + \dot{\bfy}_j F_{ki} + \dot{\bfy}_k
F_{ij} \right) \rangle 
= \langle i \frac{d}{dt}  {\rm Tr}\; \left(
\bfy_i [\bfy_j, \bfy_k] \right) \rangle \approx 0\,.
\]
Finally a term proportional to $F_{ij}F_{kl}$ has been dropped because
it vanishes by the Jacobi identity:
\[
{\rm Tr} \left(F_{ij} F_{kl} + F_{ik} F_{lj} + F_{il} F_{jk}\right) =
-{\rm Tr} \left(\bfy_i \bigl([\bfy_j,[\bfy_k,\bfy_l]] + {\rm
cyclic}\bigr) \right) = 0 \,.
\]

We can use the following relations (and their counterparts for the
$\tilde{\bfy}$ variables) to integrate by parts on the two separate
membranes (this has no effect on the time-averaged quantities).
\begin{equation}
\frac{d}{dt} {\rm Tr}\;  (\bfy_i \dot{\bfy}_i) =
{\rm Tr}\;  \left(\dot{\bfy}{}_i^2 -\bfy_i[[\bfy_i, \bfy_j], \bfy_j] \right)
= {\rm Tr}\;  \left(\dot{\bfy}{}_i^2 - F_{ij}^2 \right) \approx 0
\label{eq:parts-1}
\end{equation}
\begin{equation}
\frac{d}{dt} {\rm Tr}\;  (\bfy_i \dot{\bfy}_j) =
{\rm Tr}\;  \left(\dot{\bfy}_i\dot{\bfy}_j - \bfy_i [[\bfy_j, \bfy_k],
\bfy_k] \right) 
= {\rm Tr}\;  \left(\dot{\bfy}_i\dot{\bfy}_j + F_{ik} F_{kj} \right) \approx 0
\label{eq:parts-2}
\end{equation}
From (\ref{eq:parts-2}) we have
\[
\langle  {\rm Tr}\;\left\{
\dot{\bfy}_i\dot{\bfy}_j \dot{\tilde{\bfy}}{}_i\dot{\tilde{\bfy}}{}_j
+\dot{\bfy}_i\dot{\bfy}_j \tilde{F}{}_{ik}\tilde{F}{}_{kj}
+F_{ik}F_{kj} \dot{\tilde{\bfy}}{}_i\dot{\tilde{\bfy}}{}_j
+F_{ij}F_{jk} \tilde{F}{}_{kl}\tilde{F}{}_{li}
\right\} \rangle = 0
\]
so
\[
\langle W_{\rm cross} \rangle = {\rm Tr}\; \left\{
  48 \; \dot{\bfy}_i\dot{\bfy}_i \dot{\tilde{\bfy}}{}_j\dot{\tilde{\bfy}}{}_j
+ 24 \; \dot{\bfy}_i\dot{\bfy}_i \tilde{F}{}_{jk}\tilde{F}{}_{jk}
+ 24 \; F_{jk}F_{jk} \dot{\tilde{\bfy}}{}_i\dot{\tilde{\bfy}}{}_i 
- 12 \; F_{ij}F_{ij} \tilde{F}{}_{kl}\tilde{F}{}_{kl}
\right\}\,.
\]
Applying (\ref{eq:parts-1}) a number of times, we find
\begin{eqnarray*}
\langle W_{\rm cross} \rangle & = & \frac{7}{9} {\rm Tr}\; \left\{
  48 \; \dot{\bfy}_i\dot{\bfy}_i \dot{\tilde{\bfy}}{}_j\dot{\tilde{\bfy}}{}_j
+ 24 \; \dot{\bfy}_i\dot{\bfy}_i \tilde{F}{}_{jk}\tilde{F}{}_{jk}
+ 24 \; F_{jk}F_{jk} \dot{\tilde{\bfy}}{}_i\dot{\tilde{\bfy}}{}_i 
+ 12 \; F_{ij}F_{ij} \tilde{F}{}_{kl}\tilde{F}{}_{kl}
\right\}\\
& = & \frac{96 \cdot 14}{9} {\rm Tr}\; \left\{
\Bigl(\frac{1}{2} \dot{\bfy}^2_i + \frac{1}{4} F_{ij}^2 \Bigr)
\cdot\Bigl(\frac{1}{2} \dot{\tilde{\bfy}}^2_k
+ \frac{1}{4} \tilde{F}{}_{kl}^2 \Bigr)
\right\}.
\end{eqnarray*}
Comparing this to (\ref{eq:compare-2}) we see that the theorem has
been proven.  Thus, we have shown that any two finite size matrix
objects which have the correct leading long-range interaction with a
graviton have the correct supergravity interaction with one another.
Note that unlike the result in the previous subsection, this proof did
not depend upon the matrix-membrane correspondence and therefore is
valid for any $N$.

\section{Conclusions}

In this paper we have discussed a number of aspects of membrane
solutions of Matrix theory.  We showed that, in appropriate regimes,
finite-$N$ Matrix theory has states which behave as semiclassical
membranes.  We calculated the potential between two objects, and
established a general pair of theorems which show that in the large
$N$ limit the leading long-range force between any pair of classical
Matrix states depends only upon the energies of the states, after
radiation effects are averaged out.  This is strong evidence for the
conjecture that large-$N$ Matrix theory reproduces supergravity.  In
particular, it seems that the equivalence principle follows from
algebraic manipulations in Matrix theory.  At finite $N$, on
the other hand, we found that the Matrix potential does not agree with
the naive supergravity potential.  This suggests that the low energy
effective action of DLCQ M-theory is not given by DLCQ supergravity.

In this paper we only considered one-loop effects in Matrix theory.
Recently, a number of papers have considered two-loop effects
\cite{Becker-Becker,ggr,bbpt,ChepelevTseytlin}; it would be
interesting to see how the results in this paper are modified at
higher order.

There are a number of interesting directions in which the work in this
paper might lead.  Perhaps the most interesting questions about Matrix
membranes involve the fate of a membrane as it shrinks.  Work is in
progress to give a more detailed description of this phenomenon;
however, we give here a brief qualitative picture of the story.  A
very large spherical membrane is well described by a semiclassical
large $N$ Matrix theory configuration such as those we have considered
in this paper.  As the membrane begins to contract, it will emit
quadrupole radiation since it is invariant only under an $O(3)$
subgroup of the full $O(10)$ rotation symmetry.  It is possible to
study this radiation quantitatively in the Matrix theory picture
\cite{Dan-Wati2}.  When the membrane contracts still further, it will
eventually approach its Schwarzschild radius.  At this point, from the
supergravity point of view we would expect the membrane to become a
black hole, and to see Hawking radiation start to emerge.
Qualitatively, it is clear that this does correspond to what happens
in the Matrix picture.  The initial membrane state corresponds to a
superposition of eigenstates of the Hamiltonian for $N$ 0-branes.  If
the initial membrane configuration is sufficiently large, a
wavefunction localized around a classical configuration will have
essentially zero overlap with any state in which a majority of the
0-branes are bound together.  Thus, as the quantum state evolves over
time in Matrix theory we expect that gravitons should emerge
sporadically until the state has completely evaporated.  It would be a
significant coup for Matrix theory if it could be shown that the rate
of graviton emission from such a state corresponds to the expected
rate of Hawking radiation.  Recent works on Schwarzschild black holes
in Matrix theory
\cite{bfks,Klebanov-Susskind,Halyo-bh2,Horowitz-Martinec,Li-bh,bfksII} have
focused on verifying the energy-entropy relation using thermodynamic
or mean field arguments for Yang-Mills field theory and quantum
mechanics.  (Earlier work on non-Schwarzschild black holes appeared in
\cite{lm2,dvv-bh,Halyo-bh1}).  The states we consider here provide a natural
framework for a more detailed analysis of how these arguments work in
noncompact space.  In particular, in \cite{bfks,Klebanov-Susskind} the
thermodynamic analysis was performed by considering fluctuations
around a state described by $N$ 0-branes uniformly distributed over a
compactification torus.  The spherical state we consider, which has
0-branes uniformly distributed over its surface, forms a natural
background for an analogous calculation.  In this background, the
0-branes are ``tethered'' to the membrane; as discussed in
\cite{bfksII}, this effectively distinguishes the degrees of freedom
of the individual 0-branes, leading to the correct black hole entropy
formula.  A related picture was discussed in \cite{Horowitz-Martinec},
where it was suggested that a black hole could be formed by assembling
a large number of independent membrane configurations.  The arguments
of that paper seem to apply equally well to the situation considered
here where a single collapsing membrane forms a black hole.  Thus, at
this point it seems that there are several related arguments
indicating that the entropy-energy relations of a black hole are
correctly reproduced for a gas of 0-branes attached to a classical
configuration such as we consider here.  The outstanding challenge is
clearly to derive the correct rate of Hawking radiation from such a
state.

\section*{Acknowledgements}

We thank S.~de Alwis, T.~Banks, C.~Callan, M.~Gutperle, I.~Klebanov,
G.~Lifschytz, V.~Periwal, N.~Seiberg and L.~Thorlacius for useful
discussions.  The work of DK is supported in part by the Department of
Energy under contract DE-FG02-90ER40542 and by a Hansmann Fellowship.
The work of WT is supported in part by the National Science Foundation
(NSF) under contract PHY96-00258.

\bibliographystyle{plain}

\end{document}